\documentclass[12pt]{article}\begin{document}
\title{Representations of CCR algebras in Krein spaces of
entire functions}\author{M. Mnatsakanova,
\\ Inst. Nucl. Phys. State University,
 Moscow, Russia  \and G. Morchio, \\ Dip. di Fisica
dell'Universit\`a and INFN, Pisa, Italy \and F. Strocchi,  \\
Scuola Normale Superiore and INFN, Pisa, Italy  \and Yu. Vernov,
\\ I N R,  Russian Academy of
Sciences, Moscow, Russia}

\date{} \maketitle

\def \be {\begin{equation}}
\def \ee {\end{equation}}
\def \ume {{\scriptstyle{\frac{1}{2}}}}
\def \ra {\rightarrow}
\def \Ra {\Rightarrow}
\def \eqq {\equiv}

\def \a {{\alpha}}
\def \b {{\beta}}
\def \g {{\gamma}}
\def \d {{\delta}}
\def \eps {{\varepsilon}}
\def \th {{\theta}}
\def \l {{\lambda}}
\def \La {{\Lambda}}
\def \s {{\sigma}}
\def \Si {{\Sigma}}
\def \t {{\tau}}
\def \ph {{\varphi}}
\def \phb {{\overline{\varphi}}}
\def \o {{\omega}}
\def \Om {\mbox{${\Omega}$}}
\def \Ga {{\Gamma}}

\def \A {{\cal A}}
\def \B {{\cal B}}
\def \C {{\cal C}}
\def \D {{\cal D}}
\def \F {{\cal F}}
\def \G {{\cal G}}
\def \H {\mbox{${\cal H}$}}
\def \J {{\cal J}}
\def \K {{\cal K}}
\def \L {{\cal L}}
\def \N {{\cal N}}
\def \O {{\cal O}}
\def \P {{\cal P}}
\def \S {{\cal S}}
\def \U {{\cal U}}
\def \V {{\cal V}}
\def \W {{\cal W}}
\def \Z {{\cal Z}}

\def \Id {\mathop{\bf Id}}
\def \id {{\bf 1 }}
\def \eijk {{\varepsilon_{ijk}}}
\def \eklm {{\varepsilon_{klm}}}
\def \dij {{\delta_{ij}}}
\def \dik {{\delta_{ik}}}
\def \djk {{\delta_{jk}}}
\def \dkl {{\delta_{kl}}}
\def \Psio {{\Psi_0}}

\def \di {{\partial_i}}
\def \dj {{\partial_j}}
\def \dk {{\partial_k}}
\def \dl {{\partial_l}}
\def \do {{\partial_0}}
\def \dz {{\partial_z}}

\def \dmu {{\partial_\mu}}
\def \dnu {{\partial_\nu}}
\def \dla {{\partial_\lambda}}
\def \dr {{\partial_\rho}}
\def \ds {{\partial_\sigma}}
\def \dt {{\partial_t}}
\def \do {{\partial_0}}
\def \dum {{\partial^\mu}}
\def \dun {{\partial^\nu}}
\def \Amu {{A_\mu}}
\def \Anu {{A_\nu}}
\def \Fmn {{F_{\mu\,\nu}}}

\def \abf {{\bf a}}
\def \bbf {{\bf b}}
\def \cbf {{\bf c}}
\def \hbf {{\bf h}}
\def \k {{\bf k}}
\def \kbf {{\bf k}}
\def \jbf {{\bf j}}
\def \j   {{\bf j}}
\def \nbf {{\bf n}}
\def \q {{\bf q}}
\def \qbf {{\bf q}}
\def \p {{\bf p}}
\def \pbf {{\bf p}}
\def \sbf {{\bf s}}
\def \rbf {{\bf r}}
\def \ubf {{\bf u}}
\def \vbf {{\bf v}}
\def \xbf {{\bf x}}
\def \x {{\bf x}}
\def \y {{\bf y}}
\def \ybf {{\bf y}}
\def \v {{\bf v}}
\def \z {{\bf z}}
\def \zbf {{\bf z}}
\def \Rbf {{\bf R}}
\def \Cbf {{\bf C}}
\def \Nbf {{\bf N}}
\def \Zbf {{\bf Z}}
\def \Abf {{\bf A}}
\def \Jbf {{\bf J}}

\newcommand{\mbf}[1] {\mbox{\boldmath{$#1$}}}

\def \AO {{\cal A}({\cal O})}
\def \AO' {{\cal A}({\cal O}')}
\def \Aob {\A_{obs}}
\def \dxy {\delta(x-y)}
\def \at {{\alpha_t}}
\def \ax {{\alpha_{\x}}}
\def \atv {{\alpha_t^V}}

\def \fR {{f_R}}
\def \hf {\tilde{f}}
\def \tilf {\tilde{f}}
\def \tilg {\tilde{g}}
\def \tilh {\tilde{h}}
\def \tilF {\tilde{F}}
\def \tilJ {\tilde{J}}

\def \cc  {\subseteq}
\def \] {\supseteq}

\def \pio {{\pi_\o}}
\def \pom {{\pi_{\Omega}}}
\def \Hom { {\H_{\Omega}} }
\def \Psiom  { \Psi_{\Omega} }

\def \Pf {{\bf Proof.\,\,}}
\def \limx {{\lim_{|\x| \ra \infty}}}
\def \frx {f_R(x)}
\def \limR {\lim_{R \ra \infty}}
\def \limV {\lim_{V \ra \infty}}
\def \limko {\lim_{k \ra 0}}

\def \Roo {R \ra \infty}
\def \ko {k \ra 0}
\def \jo {j_0}
\def \su {{ \left(
\begin{array}{clcr} 0 & 1 \\1 & 0 \end{array} \right)}}
\def \sd {{ \left(
\begin{array}{clcr} 0 & -i \\i & 0 \end{array} \right)}}
\def \st {{ \left(
\begin{array}{clcr} 1 & 0 \\0 & -1 \end{array} \right)}}

\newtheorem{Theorem}{Theorem}
\newtheorem{Definition}{Definition}
\newtheorem{Proposition}{Proposition}
\newtheorem{Lemma}{Lemma}[section]

\def \Ga {{\Gamma}}
\def \tN {{N}}

\setcounter{section}{-1}

\begin{abstract}
Representations of CCR algebras in spaces of entire functions are
classified on the basis of  isomorphisms between the Heisenberg
CCR algebra $\A_H$ and
* algebras  of holomorphic operators. To each
representation of such algebras, satisfying a regularity and a
reality condition, one can associate isomorphisms and inner
products so that they become Krein  * representations of $\A_H$,
with  the gauge transformations  implemented by a continuous
$U(1)$ group of Krein space isometries. Conversely, any
holomorphic Krein representation of $\A_H$, having the gauge
transformations implemented as before and no null
subrepresentation, are shown to be contained in a direct sum of
the above representations. The analysis is extended to CCR
algebras with $[ a_i, \,a_j^* ] = \d_{i, j}\,\eta_i, \,\,\eta_i =
\pm 1$, $i = 1, ...M$, the infinite dimensional case included,
under a spectral condition for the implementers of the gauge
transformations.
\end{abstract}

\vspace{3mm}\noindent Keywords: Heisenberg algebra, Holomorphic
representations, Krein spaces

\vspace{1mm}\noindent MSC:  22E70, \, 30H05, \, 46C20, \, 81R05
\newpage
\section{Introduction}
After the Von Neumann classification of the Hilbert space
representations of the CCR algebras, a major step was the
Fock-Bargmann-Segal (FBS) analysis of the representations in
Hilbert spaces of holomorphic functions ~\cite{FBS},which has
played a crucial role for the solution of mathematical and
physical problems, especially in connection with the classical
limit and the coherent state representations ~\cite{Kl}. As it is
well known, covariant quantizations of the electromagnetic CCR
algebra, of non abelian gauge fields and of covariant strings
require representations in indefinite (Krein) spaces ~\cite{GB,
NO, D}. E.g. the commutation relations $[A^\mu(\x),
\dot{A}^\nu(\y)] = - g^{\mu\,\nu} \d(\x - \y)$ exclude a Hilbert
space realization with a Lorentz invariant vacuum. Indefinite
representations also arise in the Pauli-Villars renormalization
scheme and in general in regularization procedures ~\cite{Di}.

The analog of Von Neumann classification for the case of Krein
space representations of the Heisenberg  algebra $\A_H$, has been
discussed in ~\cite{MMSV, MS}. The aim of this note is to analyze
the Krein representations of the Heisenberg algebra on spaces of
entire functions, with the same motivations and  applications of
the FBS analysis. A complete classification of holomorphic
Hilbert-Krein
* representations will be obtained under a regularity condition
which amounts to the implementability of the group of gauge
transformations and generalizes the condition that the spectrum of
$N = a^* \, a$ is discrete.

The regularity condition leads to a representation of an extended
Heisenberg Lie algebra isomorphic to the Lie algebra $\G(0,1)$,
with discrete spectrum of one of its generators. Representations
of $\G(0, 1)$ with this property have been extensively studied in
the literature,  on (Hilbert) spaces of holomorphic functions (of
two variables), ~\cite{M} but a classification of the
* representations of the Heisenberg algebra in Krein
spaces of entire functions (of one variable) seems to be lacking.

\def \G {{\cal G}}
Our analysis is based on the following steps: \,\,{\em i)} the
characterization of the isomorphisms $\s$ between $\A_H$ and the *
algebras $\A^K$ obtained from the algebra $\A$, generated by the
holomorphic operators $z , \,\dz$, through the introduction of a
conjugation $K$, any two such isomorphisms  being related by a
transformation in a group $\G \sim SL(2, \Cbf)$ of automorphisms
of $\A$; \,\, {\em ii)} a classification of representations $\pi$
of $\A$ under the regularity condition that a $U(1)$ subgroup of
$\G$ is implementable in $\pi$; the identification of the subgroup
$\S \subset \G$, which is implementable in the space $\F$ of
entire functions leads to  a reduction of the regularity subgroups
to either the Bargmann case, namely $z \ra e^{i s} z$, $\dz \ra
e^{- i s} \dz$,  or  the Schroedinger case, $(z \pm \dz) \ra
e^{\mp i s}\,(z \pm \dz)$; \,\, { \em iii)} to each irreducible
representation of $\A$, satisfying a reality condition, one can
associate isomorphisms $\s$ and inner products $< \,, \,
>$ so that they become Krein * representations of $\A_H$ with the
gauge transformations implemented by a $U(1)$ group of Krein space
isometries $U(s)$, continuous in a Hilbert-Krein topology;
\,\,{\em iv)} conversely,  any holomorphic Krein representation of
$\A_H$, in which the gauge transformations are implemented by a
group of operators $U(s)$, continuous in $s$ in a Hilbert-Krein
topology, with no null subrepresentation, is contained in a direct
sum of the  representations obtained above. As in the Hilbert
space case, all the irreducible Krein regular representations of
$\A_H$, classified in ~\cite{MMSV, MS}, are Krein equivalent to
holomorphic Krein representations.

The above analysis, for simplicity done in the case of one degree
of freedom, can be extended to the representations of CCR algebras
$\A_H(\eta)$ for $M$ degrees of freedom, generated by elements
$a_i, \,a_i^*$, $ i = 1, ...M$, (including the case $M = \infty$)
satisfying $$[\,a_i, \,a_j \,] = 0 = [\,a_i^*, \,a_j^*\,], \,\,\,
[\, a_i, \, a_j^* \,] = \d_{i, j}\,\eta_{i}, \,\,\eta_i = \pm 1.$$
In this way one covers the field theory cases mentioned above. The
classification, discussed in Sect. 4, is obtained under a {\em
spectral condition} for the implementers of the gauge
transformations.

\section{Holomorphic Heisenberg
algebras}
We consider the vector space $\F$ of entire functions of
one variable, endowed with the standard topology $\tau$ of  the
sup over compact sets.

 The {\bf  algebra of holomorphic operators}
$\A$, is the polynomial algebra generated by $z$ and
$\partial/\partial\,z \eqq \dz$. We shall denote by $\A^K$ the *
algebra obtained by associating  to $\A$ an antilinear involution
$K$, which leaves stable the vector space $\A_1$ generated by $z$
and $\dz$. In the following, we shall use the notation ($C_K$ a $2
\times 2$ matrix) \be{\Z = \left(\begin{array}{c} z\\ \dz
\end{array} \right); \,\,\,\,K(\Z) \eqq \Z^* = \left( \begin{array}{c} z^* \\
\dz^*
\end{array} \right)  = C_K \,\Z.}\ee
The {\bf Heisenberg algebra} $\A_H$ is the polynomial *-algebra
generated by an element $a$, satisfying $ [\,a, \,a^*\,] = 1$.
\def \Vb {{\bar{V}}}
\newcommand{\SLC}{\ensuremath{SL(2, \Cbf)\,}}
\newcommand{\SLR}{\ensuremath{SL(2, \Rbf)\,}}

Special cases of isomorphisms between  the Heisenberg algebra and
the * algebras $\A^K$ are i) the {\em FBS} realization given by $
a = \dz, \,\,a^* = z,\,\,\,C_B = \s_1$, ii) the {\em Schroedinger
realization} corresponding to $a = (z + \dz)/\sqrt{2}, \,\,a^* =
(z - \dz)/\sqrt{2}, \,\,C_S = \s_3$. The isomorphisms between the
Heisenberg algebra $\A_H$ and the * algebras $\A^K$ are
characterized by
\begin{Proposition} The group $\G$ of automorphisms $\b$ of $\A$,
which leave invariant the subspace $\A_1$, is isomorphic to the
group $SL(2, \Cbf)$: \be{\b(\Z) = T \Z, \,\,\,T \in SL(2,
\,\Cbf).}\ee All the * algebras $\A^K$ are isomorphic to $\A_H$
and all isomorphisms between $\A_H$ and the * algebras $\A^K$,
mapping the linear span of $z, \, \dz$ into $\A_1$, are given by
\be{
 \left(
\begin{array}{c} a^* \\ a \end{array}\right) = V \Z,\,\,\,\,V \in SL(2, \Cbf),\,\,\,C_K =
\bar{V}^{-1}\,\sigma_1\,V,}\ee with $\s_1$ the first  Pauli matrix
(having one on the off diagonal).

The group of Bogoliubov transformations of $\A_H$ corresponds, for
each $K$, to the group $\G^K$,  isomorphic to $\SLR  \sim\, Sp(2,
\Cbf)$, of *-automorphisms of $\A^K$,  given by the matrices $T$
which satisfy $\,\, det\, T = 1,\,\,\, \bar{T} \,C = \,C \,T$.
\end{Proposition}

The analysis of the representation of $\A$ in $\F$ crucially
involves the implementation of the above automorphisms.  An
automorphism $\b \in \G$ is said {\bf implementable} in $\F$ if
there exists an operator $T_\b$ on $\F$ such that \be{ \b(\Z) \,f
= T_\b\,\Z\, T_{\b}^{-1}\,f, \,\,\,\,\forall f \in \F.}\ee A
subgroup $G \subseteq \G$ is said to be {\em represented} in $\F$
if eq.(4) holds for each $\b \in G$ and the operators $T_\b$ obey
the group law of $G$. The Lie algebra of  $\G$  is represented in
$\F$ by \be{\pi(\s_3) = z\, \dz + \ume, \,\,\,\,\pi(\s_1) =
\ume(\dz^2 - z^2), \,\,\,\,\,\pi(\s_2) = i \,\ume( \dz^2 + z^2
).}\ee

\begin{Proposition} The subgroup $\S$
of \SLC defined by \be{\s(\Z) =\left(
\begin{array}{clcr} \a  &  0 \\ \b  & \a^{-1} \end{array} \right)
\left(\begin{array}{c} z\\ \dz
\end{array} \right) \eqq S(\a,\b) \Z,  \,\,\a, \b \in \Cbf,}\ee
is represented in $\F$ in the following way \be{f(z) \ra
\Ga_S\,f(z) = f_S(z) = f(\a \,z) e^{- \a^{-1} \,\b z^2/2}.}\ee The
operators $\Ga_S$ are $\tau$-continuous; the representation is
$\tau$-differentiable in $S$ with generators in $\A$ and therefore
$\Ga_{S(\a, \b)} \,f$ is $C^\infty$ in $\a, \,\b$ in the
$\tau$-topology; the (Lie algebra) generators of the
representation of $\S$ are $z\,\dz = \pi(\s_3) - \ume$ and $-
z^2/2 = \pi(\s_1 + i \s_2)/2 \eqq \pi(\s_+)$.

Only the automorphisms belonging to $\S$ are implementable in
$\F$.
\end{Proposition} \Pf In fact, $(\s(\Z)\,f)(z) = (\Ga_S \,\Z
\,\Ga_S^{-1} f)(z)$. The identification of the (Lie algebra)
generators of $\S$ is easily obtained from the above equations,
(on $\Z$ their action  is given by the transposed matrices).

If an automorphism  $ \g \in \G, \,\g \notin \S$  is
implementable, so is the subgroup generated by $\g$ and $ \S $,
which always contains the subgroup $\b_-(s), \,\,s \in \Cbf$,
generated by $\s_- \eqq (\s_1 - i \s_2)/2$; now $\b_-(s)(z) = z +
 s \,\partial_z$ and, since, $\forall s \neq 0$, there is a $g \in \F
 $ satisfying $(z +  s\, \partial_z) g = 0$, eq.(15)
cannot hold.

\vspace{2mm}The (Lie algebra) generators of the one parameter
subgroups of $\G$ are of the form $\s \cdot n = \s_i n_i$, with
$n$ a complex vector. $\s \cdot n $ generates a two (real)
parameter subgroup $G_n$ of the form $\exp{( \l \s \cdot n)}, \,
\l \in \Cbf$ and, if $n^2 = n'\,^2, \,\, n, \,n' \neq 0$,  the two
subgroups $G_n, \,\,G_{n'}$ are \SLC conjugate.

Since $\s \cdot n$ and $\s \cdot \lambda n, \,\,0 \neq \lambda \in
\Cbf$, generate the same subgroup,  the two (real) parameter
subgroups $G_n$ fall in two equivalence classes with respect to
\SLC: those corresponding to  $n^2 =0$, called {\em degenerate},
and those with $n^2 \neq 0$. This implies that all $G_n$, with
$n^2 \neq 0$, are isomorphic to $U(1) \times R$. The other $G_n$,
with $n^2 =0$, are isomorphic to $R^2$.

The equivalence classes of the subgroups $G_n$, with respect to
$\S$, are given by the orbits under $\S$  in the adjoint
representation of \SLC. Since the generic matrix $S \in \S$ can be
written in the form $\exp{(a \s_3)}\,\exp{(b \s_+)}, \,\,a, \,b
\in \Cbf$, the (adjoint) action of $S$ on $\s \cdot n = n_3 \s_3 +
n_- \s_+ + n_+ \s_-, \,\,\,n_3,\, n_\pm \in \Cbf$, is of the form
$ (n_3, \,n_-, \,n_+) \ra (n_3 + b n_+, \, e^{2 a}(n_- -2b\,n_3 -
b^2\,n_+), \, e^{-2a}\,n_+)$. Hence, we have
\begin{Proposition} The orbits defined by $\S$ in the adjoint
representation of \SLC are the following: \vspace{1mm}\newline 1)
the orbit $\{ \s_+ \}$; it corresponds to a subgroup $G_n$ with
$n^2 = 0$; \vspace{1mm}\newline 2) the orbit $\{ \S\,\s_3 \}$; it
consists of the set $\{ \s_3  + n_- \,\s_+, \,\,n_- \in \Cbf\, \}$
\vspace{1mm}\newline 3) the orbit   $\{\S \,\s_1 \}$; it consists
of the set $ \{\, n_3 \s_3 + n_-\s_+ + n_+ \s_-, \,n_+\,\neq 0,
\newline \,\,\,\,\,n_3^2 + \, n_+\,n_- = 1 \}$, since for a
generic $S$, \,$( 0,\,1, \,1) \ra ( b,\, e^{2a}(1 - b^2),\,
e^{-2a})$ \vspace{1mm}\newline 4) the orbit $\{ \S \s_- \}$, which
corresponds to a subgroup $G_n$ with $n^2 =0$, and consists of the
set $\{ b e^{2a}\,\s_3  - (b e^{2a})^2 \s_+ + \s_- \}$.
\end{Proposition}


\section{Regular representations on entire functions}
 By
Proposition 2 the group $\S$ is implementable in $\F$, eq.(7), and
two representation are called {\em $\S$-equivalent} if the
corresponding representation spaces are related by a
transformation of $\S$ . Any representation $\pi$ in $V_{\pi}
\subseteq \F$ is automatically infinite dimensional and faithful,

In analogy with Nelson's strategy of analyzing Lie algebra
representations in terms of exponentiability of quadratic
operators ~\cite{N} we introduce
\begin{Definition} A $U(1)$ subgroup $\b^s, \,\,s  \in [0, \,2
\pi)$, of the  automorphisms $\G \simeq \SLC$,  is said to be {\bf
regularly represented} in a representation $\pi$ of $\A$ contained
in $\F$ if it is implementable in $\pi$ by a $U(1)$ group $U(s),
\,s \in [0,\,2\pi)$, $\tau$-continuous in $s$ and generated by an
element $\tN \in \A$, i.e. $\forall f \in V_\pi$ \be{\frac{d}{d s}
U(s)\,f = i \,\tN U(s) \,f , \,\,\,\tN \in \A,}\ee the derivative
being taken in the $\tau$-topology.

A representation $\pi$ of $\A$ is said to be {\bf regular} if
there are  $U(1)$ subgroups of $\G$  regularly represented in
$\pi$, called  {\bf regularity subgroups} of $\pi$.
\end{Definition}

Eq.(8) implies that actually $U(s)\,f$ is $C^\infty$ in $s$ and
therefore a regular representation is the close analog of
G\mbox{\aa}rding domain for the generators of $U(1)$ groups,
(however, in general, for fixed $s$, $U(s)$ is not a
$\tau$-continuous operator). The above regularity condition
generalizes the regularity condition on $N = a^*\,a$, which plays
a crucial r\^{o}le in the analysis of Hilbert space
representations of the Heisenberg algebra. As we shall see, such
generalized purely algebraic regularity condition leads to a
classification  of the irreducible Hilbert-Krein
* representations of the Heisenberg algebra on entire functions in
terms of those  with discrete spectrum of the $N$ operator. The
regularity condition implies that the representations of $\A$
extend to representations of the Lie algebra of the group
$G(0,1)$; by Proposition 4, their classification follows from the
analysis of those  with discrete spectrum of one of its
generators, in standard (Bargmann or Schroedinger) form, covering
in this way  a much wider class of
 representations with respect to
Ref. ~\cite{M}.

The  one parameter $U(1)$ subgroups $U(s)$ of \,\,\, \SLC are
generated by $\s \cdot n$, $n$ a unit real vector. By
 Proposition 3, $U(s)$ or $U(-s)$ is
conjugated by transformations of $\S$  either to the subgroup
$\b^s_B$  generated by $\s_3$, represented in $\F$ by $ z\,\dz +
\ume \eqq N_B + \ume$, or to the subgroup $\b^s_S$ generated by
$\s_1$, represented by $ \ume( \dz^2 - z^2) \eqq - N_S - \ume$.
Accordingly, the regularity subgroups will be called of {\em
Bargmann} or of {\em Schroedinger type}.
\begin{Proposition} Let $\pi$ be a regular representation of $\A$, $V_\pi$
its representation space and $U(s)$ the representative of
a regularity subgroup, then
\newline i) $\forall f \in V_\pi$ and $\forall k \in \Zbf$ \be{ (2
\pi)^{-1} \int_0^{2 \pi}\,d s\, U(s)\,e^{- i k s }\,f \eqq f_k}\ee
exists in the $\tau$ topology and belongs to $\F$,
\newline ii) $ f(z) = \sum_{k} \,f_k(z)$, the  series being
convergent in the $\tau$ topology, so that at least one $f_k \neq
0$; the sequence $\{f_k\}$ is of fast decrease in $k$, in all the
$\tau$-topology seminorms
\newline iii) $f_k$ is an eigenvector of the generator of $U(s)$
and, modulo $\S$ transformations, satisfy either one of the
following equations \be{ z \dz \,f_k = k \,f_k,}\ee \be{
\ume\,(-\dz^2 + z^2 - 1) \,f_k = ( k + \theta)\,f_k, \,\,\,\,
Re\,\theta \in ( - 1, \,0 ] .}\ee Given $k \in \Zbf, \,\theta \in
( -1,\, 0 ]$, eq. (11) has two non trivial solutions, given by the
parabolic cylinder functions $D_{\theta +k}(\sqrt{2}\,z)$ and
$D_{-\theta -k-1}(i \sqrt{2} z)$.
\newline iv) the
vector space $\bar{V}_\theta$, $\,\,Re \, \theta \in (- 1, 0]$
consisting of the $\tau$-convergent series $\sum c_k \,D_{\theta
+k}$ is  the carrier of a  regular representation of $\A$ with a
regularity group of Schroedinger type; conversely,  every such
representation is $\S$ equivalent to a representation contained in
$\bar{V}_\theta + S(i) \bar{V}_\theta, \,\,\,S(i) \eqq S(i, 0) \in
\S$.
\end{Proposition} \Pf. \, The existence of the integral (9)
follows from the $\tau$-continuity of $U(s)$; this also implies
that $f_k \in \F$. Point ii) follows from the standard properties
of the inversion of the Fourier series, the periodic function
$U(s) e^{i k s } f$ being $C^\infty$ in $s$. The generator of
$U(s)$  can be taken of the form \be{\tN = \sum_i n_i \,\pi(\s_i)
 - \ume - \theta \, \eqq N(n) - \theta, \,\,\,Re \,\, \theta \in
( - 1, \,0 ].}\ee In fact, since $\pi $ is faithful, the center of
$\pi(\A)$ consists of the multiples of the identity and in $\pi$
the operator $\tN - N(n)$ belongs to the center of $\pi(\A)$.

 Eqs. (8),(9) and the
$\tau$-continuity of $z$ and $\dz$ imply that $f_k$ satisfies $\tN
\,f_k = k f_k$. In the Bargmann case,  $N = z \dz - \theta$, such
an  equation has  solutions in $\F$ only for $\theta \in \Zbf$ and
therefore  $\theta = 0$. The solutions of eq.(11) are parabolic
cylinder functions ~\cite{W, M}.

By $\tau$-continuity the operators $z, \,\dz$ send
$\tau$-convergent sequences into $\tau$-convergent ones and, as we
have seen, $z \pm \dz$ act as raising/lowering operators with
respect to $k$;  therefore all the series $\sum \,c_k
\,D_{\theta+k} \,k^n, \,\,n \in \Nbf$ are $\t$-convergent so that
stability under $\A$ and regularity follows. Then, point iv)
follows from ii)-iii).

\vspace{2mm}Since $\tN_B = z \,\dz$, every Bargmann regular
representation is $\S$ equivalent to a representation invariant
under rotations. Modulo $\S$ transformations, $f_k = z^k$ and the
set of eigenvalues of $\tN_B$ is $\Nbf$.
\def \Id {\mathop{\bf Id}}
\def \id {{\bf 1 }}
\def \Flz {{F_\l(z)}}
\def \Flmz {{F_{-\l - 1}(z)}}
\def \Fnz {F_n (z)}
\def \Fliz{F_\l (i z)}
\def \Flmiz {{F_{-\l - 1} (z)}}
\def \Fniz{F_n (i z)}
\def \Fnmiz {{F_{-n - 1} (i z)}}

\def \as {{a^{*}}}
\def \Fmz {F_\mu (z)}
\def \Fmiz{F_\mu (i z)}
\def \Ae {\A_{ext}}

In the case of eq.(11), the raising and lowering  operators are $
a_+ \eqq  (z - \dz)/\sqrt{2}, \,\,a_- \eqq (z + \dz)/\sqrt{2},
\,\,[\,a_-, \, a_+\,] =1$. Putting $ \Flz \equiv  D_\l (\sqrt{2}
z)$,   with $D_\l$ the unique solution of eq.(11) with $\l =
\theta + k$, vanishing at $+\infty$ on the real line, one has
~\cite{M} \be {a_+ \Flz = F_{\l+1}(z) \ \ \ \ , \ \ \ \  a_- \,
\Flz = \l \, F_{\l-1}(z) \ \ , } \ee

If $\l$ is not an integer, the right hand sides of eqs.(13) cannot
vanish and $\Flz$ and $\Fmz$ define the same cyclic representation
space $V_\l$ iff $\l - \mu$ is an integer. The representations of
$\A$ in $ V_\l $ are irreducible, since every vector is cyclic.

For integer $\l = n$, the vector spaces $V_n$  obtained from the
cyclic vector $\Fnz$, $n \in \Zbf$, $n < 0$ define the same
representations of $\A$, which is reducible as a consequence of
eqs.(13) with integer index. In fact, $V_n$ contains the invariant
subspace $ V_0$, generated by $F_0(z)$,  actually the only non
trivial one, since in eqs.(13) only one coefficient vanishes, and
$V_0$ carries an irreducible representation of $\A$.

The other solutions of eq.(11) define $\S$ equivalent
representation spaces $V^i_{- \l - 1}$ related to $V_\l$ by the
transformation $S(i): f(z) \ra f(i z)$.

In conclusion, the representation spaces $V_S$ spanned by  the
solution of the eigenvalue equation (11) contain the following
irreducible representation:

i) $V_0$, with cyclic vector $  {F_0 (z)} $, satisfying $ a_- \,
{F_0 (z)} = 0 \ $;

ii) $V^i_{-1} = S_i \,V_0 $, with cyclic vector $ F_0 (i z) $,
satisfying $ a_+ \, F_0 (i z) = 0 \ $ ;

iii) $V_\theta$, $Re \,\theta \in (-1,0], \, \theta \neq 0$, with
cyclic vector $ F_\theta (z) $;

iv) $V^i_{- \theta - 1} = S_i\, V_\theta$, $Re\, \theta \in
(-1,0], \, \theta \neq 0$, with cyclic vector $ F_{\theta } (i z)
$.

\section{Holomorphic Krein representations}
We can now obtain a classification of *
representations of the Heisenberg algebra on inner product spaces
of entire functions.
 An inner product $< \,,\, >$ on a vector space
$V$ will be called  a {\bf Krein inner product} if it is non
degerate, it is continuous in a Hilbert space topology and the
corresponding Hilbert completion of $V$ is a Krein space, i.e. $<
f , \,g > = ( f , \, \eta\, g )$, $\,\,\eta^2 = 1$.

A {\bf Krein representation} of a * algebra $\B$ is a
representation of $\B$ on a vector space $V$ with  Krein inner
product such that the
* conjugation is represented by the inner product adjoint.
Two Krein representations $\pi_1, \,\pi_2$  on vector spaces $V_1,
\, \,V_2$ are {\bf Krein equivalent} if there is an invertible
linear transformation $U: V_1 \ra V_2$,  $U \pi_1 = \pi_2 U$,
which preserves the Krein inner product.

\begin{Definition} A {\bf holomorphic Krein representation} of $\A_H$
is a Krein representation of $\A_H$ on a space  $V$ of entire
functions, defined by one of the * isomorphisms classified in
Proposition 1 between $\A_H$ and a
* algebra $\A^K$, the algebra of holomorphic operators $z, \,
\dz$, with  conjugation $K$.

A holomorphic (Krein) representation $\pi$ of $\A_H$ is said to be
{\bf regular} if \newline i) the $U(1)$ group of gauge
transformations $\g^s, \,\,s \in [0, \,2 \pi)$ $$ \g^s( a ) = e^{-
i s } \,a,\,\,\,\,\,\g^s( a^* ) = e^{ i s } \,a^*$$ is regularly
represented (Definition 1); the implementers are generated by
$\pi(a^*\,a) + \mu $, $ \mu \in \Cbf$ and will be denoted by
$U(s)$; \newline ii) no null representation of $A_H$ is contained
in the common domain of the closures of $\pi(a), \,\pi(a^*)$.
\end{Definition}

The above notion of regularity is at the basis of the
classification of compact Lie groups in Krein spaces ~\cite{MS}.
The representation space of the above definitions is only required
to be  a dense domain in a Krein space (for Krein spaces see
~\cite{B}), since the (unbounded) representation of the Heisenberg
algebra can only be densely defined; all the operators $\pi(A),
\,A \in \A_H$ are closable since their adjoints are densely
defined. It follows from Theorem 1 below that Krein completions of
such domains can consist of holomorphic functions only for the
Fock and anti-Fock representations of the Bargmann type.
\def \Vpi {V_\pi}

>From the discussion of Sect.2 one easily gets a relation between
regular representations of $\A$ and regular representations of
$\A_H$, as an algebra. In fact, each regular representation $\pi$
of $\A$ on a space $\Vpi \subset \F$  is of the form $\pi(\A) = S
\,\pi^0(\A)\,S^{-1}$, for a suitable $S \in \S \subset SL(2, {\bf
C})$, with $\pi^0$ regular with respect to a $U(1)$ group with
implementer generated by either $N_B = z\, \dz$ or $N(\theta) =
\ume( - \dz^2 + z^2 -1) - \theta$. The representation $\pi$ gives
rise to regular representations $\pi_\s \eqq \pi \circ \s$ for
$\A_H$ as an algebra, for any isomorphism $\s \in SL(2,\Cbf)$,
$\s: \A_H \ra \A$, with \be{S^{-1}\, \s(a^*\,a - \mu )\,S   = \pm
N, \,\,\,\,N = N_B, \,\,\,\mbox{or} \,\,N = N(\theta).}\ee

We shall show that if $\pi$ is one of the irreducible
representations of $\A$ classified in Section 2, with $\theta$
real, there are Krein inner products in $\Vpi$ such that the
corresponding representations of $\A_H$ are Krein representations;
conversely, regular holomorphic Krein representations of $\A_H$
can be classified in terms of such representations.
\begin{Theorem} To each irreducible representation of $\A$, $\pi_B$ in
$V_B$, $\pi_{\theta}$   in $V_\theta$, $\theta \in (-1, 0]$, one
can respectively associate isomorphisms between $\A$ and $\A_H$,
$\s^\pm_B $, $\,\,\,\,\s^{\g, \pm}, \,\,\g \in \Rbf^+$,
(independent of $\theta$\,), and (unique) Krein inner products \
$< \,,\,>_B^\pm$, $\,\,\, < \,, \,
>^{\g,\,\pm}$ such that $ \pi \circ \s$ are holomorphic Krein
regular representations of $\A_H$, with the regularity subgroup
represented by a   group  $U(s)$ of Krein inner product
isometries, weakly continuous in $s$ with respect to a suitable
Hilbert-Krein topology.

Representations arising from different $\g$'s are Krein equivalent
and $\pi_B \circ \s_B^\pm$ are Krein equivalent to $\pi_0 \circ
\s^{\pm, \, 1}$. For each of the above $\pi$ and $\s $,  $\forall
\,S \in \S$,  $S \,\pi \circ \s \,S^{-1} $ defines a holomorphic
Krein representation of $\A_H$ in $ S\,V_{\pi \circ \s}$, with
inner product $ <  S f, \, S g > \eqq < f,\, g >_{V_{\pi \circ
\s}}$.

Conversely, for any holomorphic Krein regular representation $\pi$
of $\A_H$, with a regularity subgroup represented by $U(s)$ weakly
continuous in $s$ with respect to a Hilbert-Krein topology
$\kappa$, $V_\pi$ is contained in the $\kappa$ completion of one
of the representation spaces $ S\, V_{\pi_B \circ \s^\pm_B}$, $
S\,(V_{\pi(\theta) \circ \s} + S(i) V_{\pi(\theta) \circ \s})$, $S
\in \S$, with inner product modified in general by a $2 \times 2$
positive  matrix in the commutant of the representation .
\end{Theorem}
\Pf The construction of * representations of $\A_H$ starting from
$V_B$ and $V(\theta)$, with $\theta$ real, begins with classifying
the isomorphisms $\s$ such that \be{\s(a^*\,a) = N, \,\,\,\,N =
N_B, \,\,or \,\,\,\, N = N_S.}\ee A possible additive constant in
eq.(15) is excluded at the algebraic level, so that the
classification of the $\s$'s is independent of $\theta$; a
possible minus sign in eq. (15), see eq.(14),  only leads to a
redefinition of the inner product, leading to a new class of
representations. A "scaling" factor  $\s(a) \ra \g\,\s(a)$,
$\s(a^*) \ra \g^{-1}\,\s(a^*))$ is reduced to the case $\g > 0$ by
the implementation of the gauge transformations and in the
Bargmann case disappears by $S$-covariance. We are therefore
reduced to the following  cases.

\def \sp {\s^{1,\,+}}
\def \sgp {\s^{\g,\,+}}
\def \gp  {{\g,\,+}}
The isomorphisms $\s_B^+: \s_B^+(a) = \dz,\,\,\s_B^+(a^*) = z$,
and $\s^{1, \,+}: \sp(a) = (z + \dz)/\sqrt{2},\,\,\sp(a^*) = ( z -
\dz)/\sqrt{2}$, give rise to the standard {\em FBS} and {\em
Schroedinger} representations in  spaces of entire functions, with
the standard positive inner products $ < f, \,g
>^+_B $, $ < f,\,g >^+ $, respectively. For any $\g >
0$, one has the isomorphisms  \be{\sgp = \sp \circ \rho_\g,
\,\,\,\,\,\rho_\g(a) = \g \, a, \,\,\,\rho_\g(a^*) =
\g^{-1}\,a^*,}\ee and the corresponding inner product in $V_0$ is
given by \be{ < f, \, g >^{\gp} = < f, \, \g^{2N_S}\,g
>^+.}\ee so that  $ < f,
\,\sgp(a)\, g
>^{\gp} = < \sgp(a^*)\,f, \,g
>^{\gp}$.
All the resulting  * representations of $\A_H$ are unitarily
equivalent.

\def \Ftn {F_{\theta + n}}
\def \Ftm {F_{\theta +m}}
\def \tm {\theta + m}
\def \tn {\theta +n}
In the representation space $V(\theta)$  the inner products \be{ <
\Ftn, \, \Ftm >^{\g, +} = \d_{n, \,m} \,\g^{2n}\, \Gamma(\tm +
1),}\ee  with Hilbert majorants $ ( \Ftn, \, \Ftm )^{\g, +} =  |<
\Ftn, \, \Ftm
>^{\g, +}|$ ( $\Gamma$ the gamma function), satisfy
 $$ < \Ftn, \,\sgp(a)\, \Ftm
>^{\g, +} = \d_{n, \,m -1} \,\g^{2n +1 } (\tm) \,\Gamma(\tm )=$$
$$= < \sgp(a^*)\,\Ftn, \,\Ftm
>^{\g, +}, $$ so that one has
Krein * representations.

In all cases, the $U(1)$ group of gauge transformations is
regularly represented, by a strongly continuous unitary group,
because $\s^+_B(a^*\,a) = N_B$, $\sgp(a^*\,a) = N_S$. For fixed
$\theta$, all above representations are Krein equivalent.

Other representations are obtained by the isomorphisms \be{\s^-
\eqq \s \circ \rho^-, \,\,\,\,\rho^-(a) = a^*, \,\,\,\,\rho^-(a^*)
= - a, }\ee with $\s$ any of the isomorphisms introduced before.
In fact, in the above spaces  the inner products \be{ < f, \, g
>^- \eqq < f, \, (-1)^N\, g >^+, \,\,\,\,N = N_B, \,N_S - \theta
}\ee give rise to Krein representations. As a result, $\s_B^-$ in
$V_B$ and $\s^{1, \,-}$ in $V_0$ give rise to anti-Fock
representations of the Heisenberg algebra ~\cite{MMSV, MS}  on
entire functions, of the Bargmann and of the Schroedinger type
respectively; $\s^{1,\,-}$ in $V(\theta)$ gives rise to a
representation which is Krein equivalent to that given by $\s^{1,
\,+}$ in $V(-\theta -1)$.

Conversely, let $\pi$ be a holomorphic Krein regular
representation of $\A_H$, in $V_\pi \subset \F$ with inner product
$< \ , \ >_{\pi}$, defined by the isomorphism $\sigma : \A_H \ra
\A$.  By regularity, for some $ \mu\in \Cbf $, $ \sigma (a^* a) +
\mu \, $ generates, see eq.(8), the implementers  $U(s)$ of a
$U(1)$ subgroup of $SL(2,\Cbf)$. Therefore, by Proposition 4,
there exists $S \in \S$ such that \be {S^{-1} \sigma (a^* a + \mu)
S \eqq \rho_S^{-1} \circ \s (a^* \,a + \mu) = \pm N }\ee (the
generator of a one parameter subgroup being determined up to a
sign), with $N = N_B$ or $N = N(\theta)$. The space $V_{\pi^0}
\equiv S^{-1} V_\pi $, with the inner product $ <f \, , \, g
>_{\pi^0} \equiv < S f \, , \, S g
>_{\pi} $, carries a holomorphic Krein regular representation $
\pi^0 (A) f \equiv S^{-1} \pi (A) S f $, $ f \in V_{\pi^0} $,
defined by the isomorphism $\sigma^0 = \rho_{S}^{-1} \circ
\sigma$. Eq.(14) reads \be {\sigma^0 (a^* a + \mu)   = \pm  z \,
\dz \ \ \ \ {\rm or}\ \ \ \ \sigma^0 (a^* a + \mu)   = \ume ( (
\dz^2 - z^2 - 1 ) + \theta) . } \ee $\sigma^0$ is defined, as in
Proposition 1, by a matrix $V^{-1} \in SL(2,\Cbf)$ and this
immediately implies that eq.(22) only has  the solutions $\sigma^0
= \sigma_B^\pm \circ \rho_\gamma$,
       $\sigma^0  = \sigma_S^\pm \circ \rho_\gamma,$
respectively in the Bargmann and Schroedinger case, with $\mu =0$
in the Bargmann case, $\mu = \theta$ in the Schroedinger case.

In the Bargmann case, the automorphisms of $\A$, $ \,
\rho^B_\gamma \equiv \sigma_B^\pm \circ \rho_\gamma \circ
(\sigma_B^\pm)^{-1}$, are implemented by transformations in $\S$
and therefore \be{  \sigma_B^\pm \circ \rho_\gamma (A)  =
\rho^B_\gamma \circ \sigma_B^\pm = S_\gamma^{-1} \sigma_B^\pm (A)
S_\gamma,}\ee so that one is reduced, by a transformations in
$\S$, to $ \sigma_B^\pm $. In the Schroedinger case, for $\gamma =
\exp{-is}$, the corresponding automorphisms $\rho^S_\gamma$ are
implemented by $U_0(s) \equiv S^{-1} U(s) S$, and therefore one
has  to consider only the case $\gamma > 0$.

Representations of $\A_H$ defined by $\sigma_B^\pm, \sigma_S^\pm
\circ \rho_\gamma \ , \gamma > 0 $ will now be reduced to the
representations introduced above on spaces spanned by eigenvectors
of $N_B$ and $N_S$. Weak continuity of $U(s)$ in a Hilbert Krein
topology $\kappa$ is equivalent to weak continuity of $U_0(s)$ in
$V_{\pi^0} $, in a Hilbert-Krein topology $\kappa^0$, defined from
$\kappa$ by the isometry $S$. This implies that the integrals
defining eigenvectors, eq.(9), exist also as weak limit in the
corresponding Krein space, and that the eigenvector expansion
converges weakly in the same space. By $\tau$ continuity of $z$
and $\dz$ and their closability in the weak Hilbert-Krein
topology, one obtains a Krein representation of $\A_H$ on spaces
spanned by eigenvalues of $N_B$, or of $N_S$.

In the Bargmann case, eigenvalues are not degenerate; therefore,
given the isomorphisms $\sigma^\pm$, there is only one (up to a
constant) inner product, in the space spanned by the eigenvectors,
which gives a Krein representation. By weak convergence of the
eigenfunction expansion, the space $V_{\pi^0}$ is a subspace of a
Krein completion of the space spanned by the eigenvectors, and the
result follows since $S$ extends to a Krein space isometry.

In the Schroedinger case, the  eigenvalues of $N_S$ are real,
since $$ (\lambda + n) <f_\lambda, g_{\lambda+n}> =
   < \sigma(a^* a) f_\lambda, g_{\lambda+n}> = \overline{\lambda}
   <f_\lambda, g_{\lambda + n}> $$
for $Im\, \l \neq 0$ implies  the vanishing of the inner product.
The parameter $\theta$, eq. (12), is therefore real. For $\theta
\neq 0$, one has a Krein representation on the space spanned by
the eigenvalues, which is reducible into two irreducible
equivalent representations, so that the Krein inner product is
determined, and coincides with the inner product introduced above,
apart from an hermitean $2 \times 2$ matrix $M$ in the commutant
of the representation; $M$ must be positive (or negative)
definite, since otherwise null subrepresentations would appear.
For $\theta = 0$, the eigenvectors are $F_n(z)$, $F_n(iz)$, $n \in
\Zbf$, and only $n \geq 0$ is admitted; otherwise
 $$<F_0(z), F_0 (z)> = <a_+ F_{-1}, F_0> =  <F_{-1}, a_- F_0> = 0   $$
which implies $<F_n(z), F_m (z)>  = 0  \ \ \forall n,m \geq 0$,
i.e. a null subrepresentation in the closure of $\pi$. The same
applies to $F_n(iz)$, so that all eigenvectors belong to $ V_0 +
S(i)V_0$; therefore, the representation is contained in a Krein
completion of the unique (up to two irrelevant constants in the
inner products) representation in $ V_0 + S(i)V_0$. For all
$\theta \in (-1, 0]$, the representation is  reconstructed by weak
convergence of the eigenfunction expansion and by the fact that
$S$ extends to a Krein space isometry.

\sloppy
\section{Krein representations of infinite dimensional CCR algebras}
\fussy In this Section we extend our analysis to representations
of CCR algebras for $M$ degrees of freedom including the infinite
dimensional case.
\def \AHE {\A_H(\eta)}
\def \AM  {\A_{H, \,M}}

\def \eti {\eta_{i}}
\def \etij {\eta_{i \,j}}
The CCR algebras which arise in physically interesting models, in
particular in the Gupta-Bleuler quantization of the
electromagnetic field ~\cite{GB}, are generated by elements $a_i,
\,\,a^*_i, \,$ $i = 1, ...M$, satisfying in general commutation
relations of the form \be{[\,a_i, \,a_j \,] = 0  = [\,a_i^*,
\,a_j^*\,],\,\,\, [\, a_i, \, a_j^* \,] = \eta_{i\,j},}\ee with
$\eta_{i \,j}$ a non degenerate complex matrix. Thus, by a linear
transformation one can reduce to the case \be{ [ a_i, \,a_j] = 0 =
[ a^*_i, \,a^*_j],\,\,\, [\,a_i, \,a_j^*\,] = \d_{i, \,j}\,\eti,
\,\,\,\,\,\eti = \pm 1.}\ee The algebra $\A_H(\eta)$ generated by
$a_i, \,a_i^*,\,$ $i = 1, ...M$, (possibly $M = \infty$),
satisfying eqs.(25), is isomorphic to the Heisenberg algebra $\AM
$, for $M$ degrees of freedom (corresponding to $\eti = 1, \,i =
1, ...M$), with isomorphism given by $ \rho(a_i) = \ume (1 +
\eti)\,a_i + \ume (1 - \eti) \,a_i^*$. However, the isomorphism
does not commute with the gauge transformations \be{ \g^s(a_i) =
e^{- i s}\,a_i, \,\,\,\,\,s \in [\,0, \, 2 \pi), \,\,\,\,i = 1,
...M}\ee and therefore a classification of representations in
terms of conditions on the gauge generators, $ N = \sum_{i = 1}^M
\eti\, a^*_i\,a_i$, depends on $\eti$. In general the analysis
depends on the specification of $M$ regularity $U(1)$ groups,
leading to products of representations classified in Theorem 1.

A substantial simplification occurs if one requires positivity of
the generator of gauge transformations, a property that in the
physical applications is closely related to the energy spectral
condition, that is stability. In fact, this condition leads to a
unique Krein representation of the algebra defined by eqs.(25), up
to multiplicities, which is a Hilbert space representation iff the
spectrum of $\etij$ is positive. This result covers  the infinite
dimensional case, for algebras defined by eqs.(25).

The regularity property of Definition 2 is easily adapted to the
$M$-dimen\-si\-o\-nal case. The implementers $U(s)$ are said to
satisfy the {\bf spectral condition} if , $\forall f,\, g, \in
V_{\pi}$, $< f,\, U(s)\,g >$ extends to a bounded analytic
function in the upper half (complex) plane.
\begin{Theorem} A regular Krein representation of $\AHE$,
admitting implement\-ers $U(s)$ satisfying the spectral condition
is contained in a direct sum of representations of the form $\pi =
\pi_H \circ \rho$, where $\pi_H$ is the cyclic subrepresentation
of $\AM$ in the tensor product of Fock and antiFock
representations of $\A_H^{(i)}$, defined by a cyclic vector
$\Psio$, which, for each $i$, defines a Fock/antiFock
representation of $\A_H^{(i)}$, for $\eta_i = \pm 1$.

\def \dzi {\partial_{z_i}}
The representation is Krein equivalent to a holomorphic
representation of the Bargmann form \be{\pi(a_i) =
\dzi,\,\,\,\pi(a^*_i) = z_i}\ee on the space of polynomials in $M$
(possibly $M = \infty$) complex variables, with inner product
given for $n_i, \,m_j \geq 0$ by \be{ < \,z_{i_1}^{n_1}
...z_{i_k}^{n_k}, \,z_{i_1}^{m_1}...z_{i_k}^{m_k}\,> = \prod_{i =
1}^k  n_i! \,(-1)^{n_i(1 - \eti)/2} \d_{n_i,\,m_i},}\ee or of
Schroedinger form \be{\pi(a_i) = \sqrt{\ume}( z_i + \dzi),
\,\,\,\,\pi(a_i^*) = \sqrt{\ume}(z_i - \dzi)}\ee on the space of
polynomials in $M$ (possibly $M = \infty$) variables, multiplied
by $ \Psio(z_1, ...z_M) = \prod_{i= 1}^M e^{-z_i^2/2}$, with
scalar product given similarly to eq.(28), in terms of Hermite
polynomials
\end{Theorem}
\Pf  Given $f \in V_\pi$, by the regularity property, eq.(8)
defines, as in Theorem 1, at least one  non zero element $f_k$ of
the Krein closure of $V_\pi$, which is in the domain of the
closures of all the elements $\pi(A), \, A \in \AHE$. On the other
hand, the spectral condition implies that $\forall g \in V_\pi$
$$< g, \, U(t)\,f > = \sum_k < g, \, f_k > e^{i k t}$$ is the
Fourier-Laplace transform of a distribution $F(\o)$ with
$supp\,\,\,F(\o) \subseteq \Rbf^+$, so that  $< g, \, f_k > \neq
0$ only for $k \geq 0$ and therefore $f_k = 0$ for $k < 0$. Hence,
by closability of $\pi(A), \,\,\,A \in \AHE$, $\forall n > k$, $$
(2 \pi)\, \pi(a_{i_1}...a_{i_n}) f_k = \int_0 ^{2 \pi} d s\,e^{-i
k s} \pi(a_{i_1}...a_{i_n}) \,U(s)\,f =$$ $$ \int_0^{2 \pi} d s
\,e^{-i(k - n) s} U(s) \,\pi(a_{i_1}...a_{i_n})\,f = (2 \pi)
\,(\pi(a_{i_1}...a_{i_n})\,f)_{k - n} = 0.$$ If $\pi(a_i) \,f_k =
0, \,\,\forall i$, we consider the cyclic representation of $f_k$;
otherwise for some $i_1$, $\pi(a_{i_1}) \,f_k = (\pi(a_{i_1}\,f
)_{k - 1} \neq 0$. Within $n$ steps, we obtain a non zero vector
$\Psio$ such that $\pi(a_i) \Psio = 0, \,\forall i$. Therefore
$\pi_H \eqq (\pi \circ \rho^{-1})$ is a representation of $\AM$
with cyclic vector $\Psio$ satisfying the Fock/antiFock condition
$$\rho(a_i) \,\Psio = 0 , \,\,\mbox{if} \,\,\,\eti = 1,
\,\,\,\rho(a^*_i) \,\Psio = 0, \,\,\,\mbox{if} \,\,\,\,\eti =
-1.$$

By the same argument as in the proof of Theorem 1, the inner
product in the cyclic space is then unique up to a factor, which
cannot vanish, since otherwise one would get a null
subrepresentation. The isomorphism with the holomorphic
representations characterized in Theorem 1 is straightforward.

\end{document}